# Accessing the Host Galaxies of Long Gamma-Ray Bursts with Next-generation Telescopes

Guang-Xuan Lan[1], Ye Li[2], and Zhuo Li[1,3]
[1] Department of Astronomy, School of Physics, Peking University, Beijing 100871, People's Republic of China; gxlan@pku.edu.cn, zhuo.li@pku.edu.cn
[2] Purple Mountain Observatory, Chinese Academy of Sciences, Nanjing 210023, People's Republic of China; yeli@pmo.ac.cn
[3] Kavli Institute for Astronomy and Astrophysics, Peking University, Beijing 100871, People's Republic of China

## Abstract

We present a method to estimate the detection expectations of host galaxies of long gamma-ray bursts (LGRBs) in the $grizJHKL$ bands. It is found that given the same limiting magnitude $m_{grizJHKL,\rm lim}$ in each band, the $z$ band produces the largest number of overall LGRB hosts and low-mass hosts ($M_* \leqslant 10^8 M_\odot$) at $m_{grizJHKL,\rm lim} \gtrsim 26$ mag. For the detection of high-redshift LGRB hosts (redshift $\geqslant 5$), it is recommended to prioritize the $L$ band, due to its good performance at both low and high limiting magnitudes. We specifically estimate the expectation of LGRB host detection with $m_{grizJHKL,\rm lim} = 28$ mag, which the James Webb Space Telescope can partially attain. We find that there may exist 116, 259, 277, 439, 266, 294, 274, and 316 LGRB hosts, including 0.54, 31, 28, 143, 12, 20, 14, and 35 low-mass ones in the $grizJHKL$ bands and 13, 14, 15, 14, and 15 high-redshift ones in the $zJHKL$ bands, for 15 yr Swift LGRBs with $S \geqslant 10^{-6}$ erg cm$^{-2}$. The results show that the study of LGRB hosts under next-generation observational conditions holds significant potential, especially for low-mass host studies. However, it appears that the deeper sensitivities of galaxy telescopes may not significantly enhance statistical studies of high-redshift hosts. Strategies aimed at increasing the number of distant LGRB hosts may require the expansion of high-redshift LGRB detections.

*Unified Astronomy Thesaurus concepts:* Gamma-ray bursts (629); Galaxies (573); Stellar mass functions (1612)

## 1. Introduction

Gamma-ray bursts (GRBs) are extremely luminous explosions in the Universe. They are divided into two groups based on the bimodal distribution of their duration $T_{90}$ (Kouveliotou et al. 1993). Long GRBs (LGRBs) have a duration of $T_{90} > 2$ s and are usually supposed to be consistent with the collapse of massive stars (Woosley & Bloom 2006). On the other side, with $T_{90} \leqslant 2$ s, short GRBs (SGRBs) are believed to come from the merger of compact objects (Abbott et al. 2017; Goldstein et al. 2017). However, the boundaries of the physical origins of GRBs are actually ambiguous. The durations of some bursts (GRB 060614, GRB 211211A, and GRB 230307A) are significantly long, whereas other properties are similar to typical short bursts and are accompanied with merger phenomena of compact stars (Gehrels et al. 2006; Rastinejad et al. 2022; Troja et al. 2022; Yang et al. 2022; Levan et al. 2023; Sun et al. 2023). Similarly, some SGRBs (GRB 090426 and GRB 200826A) show afterglows and locations akin to being supernova-associated (Antonelli et al. 2009; Levesque et al. 2010a; Ahumada et al. 2021; Zhang et al. 2021; Rossi et al. 2022).

The host galaxies of GRBs are of crucial importance to identifying their potential progenitors. Normally, LGRBs tend to reside in star-forming galaxies with low luminosity, metallicity, and stellar mass (Fruchter et al. 2006; Stanek et al. 2006; Modjaz et al. 2008; Savaglio et al. 2009; Levesque et al. 2010b; Svensson et al. 2010; Graham & Fruchter 2013; Kelly et al. 2014; Vergani et al. 2015). Within the hosts, they prefer the bright regions where the offsets from the centers of galaxies are small (Bloom et al. 2002; Blanchard et al. 2016; Japelj et al. 2018). Meanwhile, SGRBs are typically associated with old star populations and are much more distant from their host centers (Berger 2009, 2014; Fong & Berger 2013), corresponding to the theoretical locations of compact binaries. The anomalous GRBs mentioned above, however, show hosts and positions unusual in the duration catalogs to which they belong. This strongly encourages one to observe GRB host galaxies to better understand the origins of GRBs.

Furthermore, investigating LGRB hosts is beneficial to revealing the evolution of LGRBs. As bright and distant (up to $z \sim 9$ currently; Cucchiara et al. 2011) products from probably the death of massive stars, LGRBs have long been expected to be used to probe the cosmic star formation rate (cSFR) at early epochs (Salvaterra 2015). However, numerous previous works have shown that redshift evolution is somehow necessary in a cSFR-based model to reproduce the redshift distribution of LGRBs (Lloyd-Ronning et al. 2002; Firmani et al. 2004; Le & Dermer 2007; Salvaterra et al. 2009b, 2012; Wanderman & Piran 2010; Wei et al. 2014; Pescalli et al. 2016; Wei & Wu 2017; Lan et al. 2019, 2021; Ghirlanda & Salvaterra 2022). The evolution of some properties of LGRB hosts, e.g., metallicity, stellar mass, and initial mass function, may be the key to explaining the evolution of LGRBs (Langer & Norman 2006; Li 2008; Wang & Dai 2011, 2014; Robertson & Ellis 2012; Perley et al. 2016b; Lan et al. 2022).

On the other hand, LGRB hosts can also provide valuable insights into the nature of galaxies. Actually, low-mass galaxies are easily overlooked in a shallow survey, due to their faint luminosity. LGRBs commonly connect with these types of galaxies (Savaglio et al. 2009), making them effective tools to pinpoint and probe such low-mass galaxies (Savaglio et al. 2009; Trenti et al. 2012; Kelly et al. 2013; Greiner et al. 2015;

Schulze et al. 2015). In investigations of distant galaxies, pushing the instruments to their limits is vitally important. As the redshifts of LGRB hosts can be measured independently via LGRB afterglows, it is feasible for LGRBs to constrain the properties of host galaxies that are even undetected (Basa et al. 2012; Tanvir et al. 2012), providing a unique tool to study galaxies in the early Universe. For high-redshift hosts that are detectable, techniques of using LGRB afterglows to characterize the host properties, such as the interstellar medium, dust extinction, chemical abundances, and so on (see Salvaterra 2015 and references therein), have also been widely exploited. There is no doubt that the observation of GRB hosts is helpful both for studying GRBs themselves and for shedding more light on galaxy nature.

However, the current sample size of GRB hosts is still too small to extensively carry out statistical research in subdivided directions. For instance, only a small number of low-mass or high-redshift GRB hosts have been characterized (Savaglio et al. 2009; Hjorth et al. 2012; Li & Zhang 2016; Perley et al. 2016b). The existing LGRB host surveys also describe the redshift evolution of LGRB formation environments inadequately (Krühler et al. 2015; Perley et al. 2016a, 2016b; Lan et al. 2022). Next-generation galaxy telescopes, e.g., the James Webb Space Telescope (JWST), the China Space Station Telescope (CSST), and so on, may open up greater opportunities to enlarge the GRB host sample. Predicting the possible number of GRB hosts that could be detected by future galaxy telescopes will contribute to the design of GRB host studies. The scope of this paper is exactly to estimate the expected number of LGRB hosts under next-generation observational conditions based on the stellar mass function of LGRB hosts. To allow the optimization of observational strategies for different science goals, multiband predictions from the optical to the mid-IR are included in this work.

This paper is arranged as follows: the methods and models used to predict LGRB host detection are briefly described in Section 2. Then the results of our predictions are presented in Section 3. In Section 4, we will discuss a few uncertainties in our predictions. Finally, a summary of this work is drawn in Section 5. Throughout this paper, a flat ΛCDM cosmological model with $H_0 = 70$ km s$^{-1}$ Mpc$^{-1}$, $\Omega_m = 0.3$, and $\Omega_\Lambda = 0.7$ is adopted.

## 2. Methods and Models

In principle, the expected number of LGRB hosts in observation can be calculated through the following formula (Ajello et al. 2009, 2012; Abdo et al. 2010; Salvaterra et al. 2012; Zeng et al. 2014, 2016; Vergani et al. 2015; Lan et al. 2019, 2021, 2022):

$$N_{\mathrm{exp,host}} = \frac{\Delta\Omega T}{4\pi} \int_0^{z_{\max}} f \cdot \frac{\psi(z)}{1+z} \frac{dV(z)}{dz} \cdot F_E(z) \cdot F_M(z) dz$$

$$F_E(z) = \int_{\max[E_{\min}, E_{\lim}(z)]}^{E_{\max}} \phi(E_{\mathrm{iso}}) dE_{\mathrm{iso}}$$

$$F_M(z) = \int_{\max[M_{*,\min}, M_{*,\lim}(z)]}^{M_{*,\max}} \varphi(M_*, z) dM_*,$$

(1)

where $\Delta\Omega$ is the field of view of the GRB detector, $T$ is its working time, and $\psi(z)$, $\phi(E_{\mathrm{iso}})$, and $\varphi(M_*)$ are the event rate density, isotropic-equivalent energy function, and host galaxy stellar mass function of LGRBs, respectively. $dV(z)/dz = 4\pi c D_L^2(z)/[H_0(1+z)^2\sqrt{\Omega_m(1+z)^3 + \Omega_\Lambda}]$ is the comoving volume element in a flat ΛCDM model, and $D_L(z)$ is the luminosity distance at redshift $z$. Finally, $f$ denotes the fraction of gamma-ray-detected LGRBs with good localization.

Based on the Swift Gamma-Ray Burst Host Galaxy Legacy Survey (SHOALS; Perley et al. 2016a, 2016b), which combines the observations of Swift ($\Delta\Omega = 1.33$ sr) and Spitzer/IRAC (3.6 $\mu$m), Lan et al. (2022) have quantified the rate, energy, and host stellar mass functions of the LGRBs mentioned above. SHOALS was selected in an unbiased manner from 7 yr Swift LGRBs and finally included 119 LGRBs with redshift completeness up to ∼92%. Using Spitzer/IRAC observations, Perley et al. (2016b) presented near-IR luminosities and stellar masses of the host galaxies of SHOALS LGRBs with limiting magnitude $m_{3.6\,\mu\mathrm{m,lim}} \sim 25$ mag, so the properties of ∼two-thirds of the hosts with measured redshift in SHOALS have been well characterized. To estimate the future detection of LGRB hosts without bias, we base our prediction on the observational results of the host galaxies in SHOALS, assuming the GRB localization capability of the future will be similar to that of the present. Although this will make our predicted results conservative, the completeness of the LGRB afterglow/host detection is secure in our calculation. Then the fraction of LGRBs with good localization $f$ can be calculated as $N^{\mathrm{SHOALS}}_{\mathrm{predicted}}(S \geqslant S_{\mathrm{lim}})/N^{\mathrm{Swift}}_{\mathrm{observed}}(S \geqslant S_{\mathrm{lim}})$, where the numerator is the predicted number of SHOALS-selected LGRBs without the stellar mass limit of the hosts, and the denominator is the observed number of Swift LGRBs with $S \geqslant S_{\mathrm{lim}}$ during the observational time of the SHOALS sample.

As the mass evolution model introduced in Lan et al. (2022) is one of the well-reproduced models of the SHOALS observations, in this work, we attempt to adopt the mass evolution model to derive the expected detection of LGRB hosts by assuming the sensitivities of galaxy telescopes in different bands. In the mass evolution model, the LGRB rate is assumed to be proportional to the cSFR as $\psi(z) = \eta\psi_\star(z)$, where $\eta = 10^{-7.67}/f\ M_\odot^{-1}$ is the LGRB formation efficiency, and the comoving cSFR (in units of $M_\odot$ Mpc$^{-3}$ yr$^{-1}$) was modeled as (Li 2008):

$$\psi_\star(z) = \frac{0.0157 + 0.118z}{1 + (z/3.23)^{4.66}}. \tag{2}$$

The LGRB energy function is considered to be a broken power-law form:

$$\phi(E_{\mathrm{iso}}) = \frac{A}{\ln(10)E_{\mathrm{iso}}} \begin{cases} \left(\dfrac{E_{\mathrm{iso}}}{E_b}\right)^a & E_{\mathrm{iso}} \leqslant E_b \\ \left(\dfrac{E_{\mathrm{iso}}}{E_b}\right)^b & E_{\mathrm{iso}} > E_b, \end{cases} \tag{3}$$

where $A = 0.10$ is a normalization factor and $a = -0.39^{+0.18}_{-0.12}$, $b = -1.23^{+0.17}_{-1.17}$, and $E_b = 53.51^{+0.54}_{-0.25}$ are the power-law indices and the break energy, respectively. The LGRBs in SHOALS were limited by fluence $S_{\mathrm{lim}} = 10^{-6}$ erg cm$^{-2}$, which is much brighter than the sensitivity of Swift, to enhance the completeness of the subsample of LGRBs (Salvaterra et al. 2012; Perley et al. 2016a, 2016b; Pescalli et al. 2016). As a result, the isotropic-equivalent energy $E_{\rm iso}$ of the surviving LGRBs is high and ranges from $E_{\min} = 10^{51}$ erg to $E_{\max} = 10^{56}$ erg. To be able to make an unbiased prediction of the LGRB host detection, we also adopt the fluence limit $S_{\lim} = 10^{-6}$ erg cm$^{-2}$ in this work in the calculation of the limiting energy of LGRBs:

$$E_{\lim}(z) = \frac{4\pi D_L^2(z) S_{\lim}}{1+z} \frac{\int_{1/(1+z)\,\rm keV}^{10^4/(1+z)\,\rm keV} EN(E)dE}{\int_{15\,\rm keV}^{150\,\rm keV} EN(E)dE}, \quad (4)$$

where 15–150 keV is the observed energy range of Swift/Burst Alert Telescope and $N(E)$ is the GRB photon spectrum. Typical values of a low-energy spectral index $-1$ and a high-energy spectral index $-2.3$ in the Band function are employed to describe the burst spectrum (Band et al. 1993; Preece et al. 2000; Kaneko et al. 2006). For a given isotropic-equivalent energy $E_{\rm iso}$, the peak energy of the spectrum $E_p$ is estimated through the empirical $E_p$–$E_{\rm iso}$ correlation (Amati et al. 2002; Nava et al. 2012), i.e., $\log[E_p(1+z)] = -29.60 + 0.61 \log E_{\rm iso}$.

The stellar mass function of LGRB hosts is assumed as an evolving Schechter form (Drory & Alvarez 2008; Lan et al. 2022) and is expressed as:

$$\varphi(M_*, z) = \varphi_*(z)\left(\frac{M_*}{M_b(z)}\right)^\xi \exp\left(-\frac{M_*}{M_b(z)}\right)\frac{1}{M_b(z)}$$
$$\varphi_*(z) = \varphi_0(1+z)^\delta$$
$$\log M_b(z) = \log M_{b,0} + \gamma \ln(1+z), \quad (5)$$

where $\varphi_0 = 0.0036$ is a normalization factor, the power-law index $\xi = -1.14^{+0.10}_{-0.06}$, the break mass $M_{b,0} = 11.24^{+0.25}_{-0.43}$, and two additional evolution parameters $\delta = 1.96^{+0.44}_{-0.47}$ and $\gamma = -0.38^{+0.28}_{-0.19}$ were estimated based on the SHOALS sample. The minimum and maximum stellar mass are empirically assumed as $M_{*,\min} = 10^7 M_\odot$ and $M_{*,\max} = 10^{13} M_\odot$, respectively, in this work.

The lower limit of the stellar mass $M_{*,\lim}(z)$ in Equation (1) depends on the sensitivities of galaxy telescopes. Its precise value should be calculated via the LGRB host spectral energy distribution (SED) template, which, however, has not yet been fully understood. Usually, apart from spectroscopy, the stellar mass of a galaxy can also be derived by IR photometry. In order to take into account the fluctuation of the LGRB host SED in our computation, the stellar mass–luminosity correlations of LGRB hosts in different bands are employed in this work to statistically estimate the stellar mass limits $M_{*,\lim}(z)$ in corresponding bands. Based on the database of GRB Host Studies (GHostS),[4] we simply investigate the stellar mass–luminosity correlations ($M_*$–$M_\lambda(z)$, where $M_\lambda(z)$ is the absolute magnitude of the LGRB hosts measured in band $\lambda$) of LGRB hosts in the $grizJHKL$ bands and show the results in the Appendix. Given a magnitude limit $m_{\lambda,\lim}$ in band $\lambda$, the stellar mass limit $M_{*,\lim}(z)$ observed in this band can be straightforwardly obtained through the correlation $M_{*,\lim}(z)$–$M_{\lambda,\lim}(z)$. Next, considering the limiting magnitudes of the JWST F090W, F115W, F162M, F210M, and F360M filters can be greater than 28 mag,[5] our predictions will start with the limiting magnitude $m_{\lambda,\lim} = 28$ mag in each band.

---
[4] http://www.grbhosts.org/

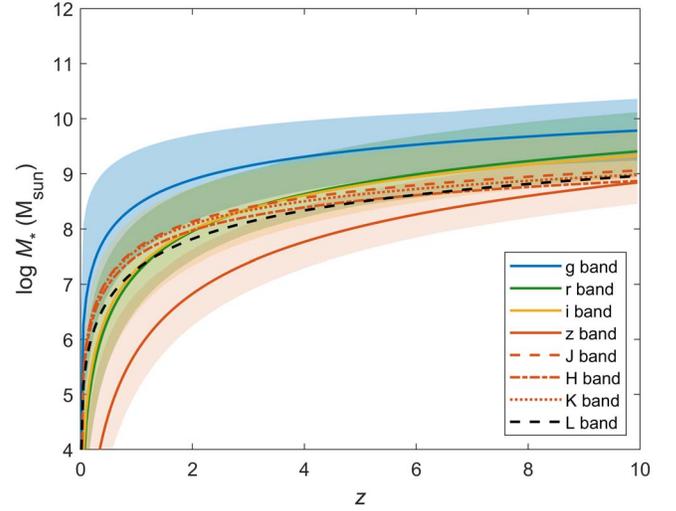

**Figure 1.** Stellar mass limits calculated via $M_*$–$M_\lambda(z)$ correlations in the $grizJHKL$ bands, where the limiting magnitude is 28 mag in each band. The results in the $griz$ bands are shown as blue, green, orange, and red lines with 1$\sigma$ confidence regions, respectively. And the results in the $JHKL$ bands are shown as red dashed, dotted–dashed, dotted, and black dashed lines, respectively. Host galaxies with stellar mass lower than these lines may suffer selection effects.

Since emission light with wavelengths shorter than the Ly$\alpha$ line may be absorbed by the intergalactic medium (IGM), we assume the maximum redshifts $z_{\max}$ of the LGRB hosts that could be detected in the $griz$ bands are 3, 4, 5.5, and 6.5, respectively. Note that the measured redshifts of LGRBs are smaller than 10 to date. Therefore, the maximum redshift $z_{\max} = 10$ is considered for the prediction in the $JHKL$ bands.

### 3. Results

In fact, the number of LGRB hosts we can detect strongly depends on the sensitivities of the GRB detectors and the galaxy telescopes. The priority of this work is to study the capacity of future galaxy telescopes for LGRB host detection. Since the historical Swift data can supply plenty of LGRB positions and many of their hosts have not yet been detected in particular, the Swift-detected LGRBs are mainly considered in our predictions. We first assume the limiting magnitude of the galaxy telescopes is 28 mag in any of the $grizJHKL$ bands to estimate the detection of LGRB hosts for different science purposes (i.e., overall, low-mass, and high-redshift host studies). Using the $M_*$–$M_\lambda(z)$ correlations of LGRB hosts (see the Appendix), we derive the stellar mass limits $M_{*,\lim}(z)$ of the LGRB host detection in the $grizJHKL$ bands, which are shown in Figure 1.

For the results from the best-fitting models, the selection effects on stellar mass exhibit similarity in different IR bands, whereas they significantly differ in optical bands. At low and middle redshifts, the stellar mass limit in the $z$ band is much smaller than that of the other bands. Even at redshift ~5, host galaxies with $M_* \sim 10^8 M_\odot$ may still be observed in this band. However, in the $g$ band, the stellar mass limit grows quickly at very low redshift, making the detection of host galaxies with

---
[5] https://jwst-docs.stsci.edu/jwst-near-infrared-camera#JWSTNearInfraredCamera-Filters

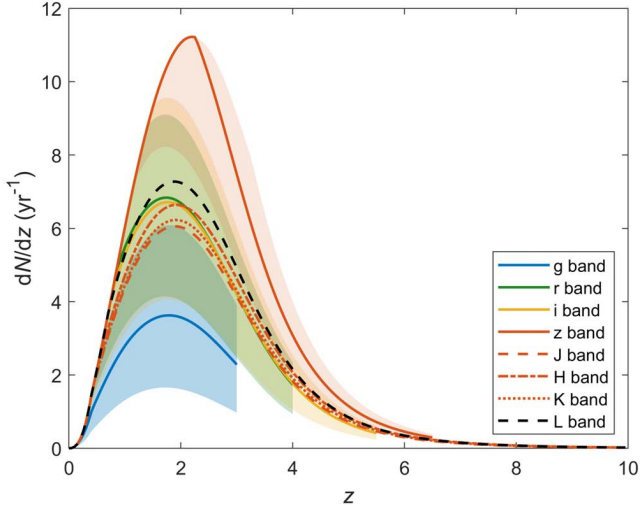

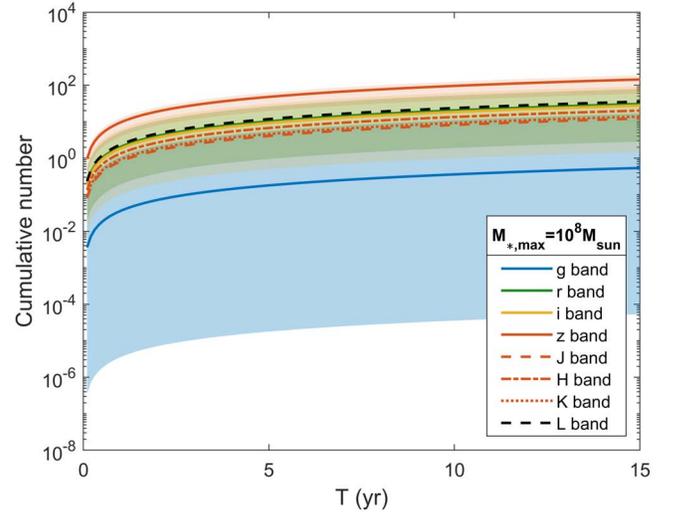

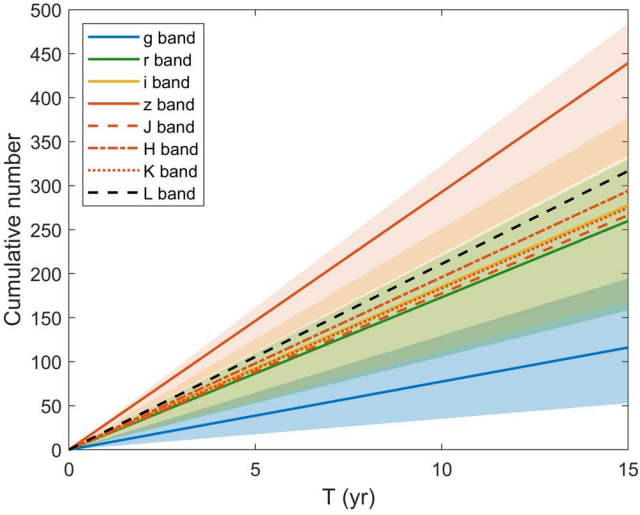

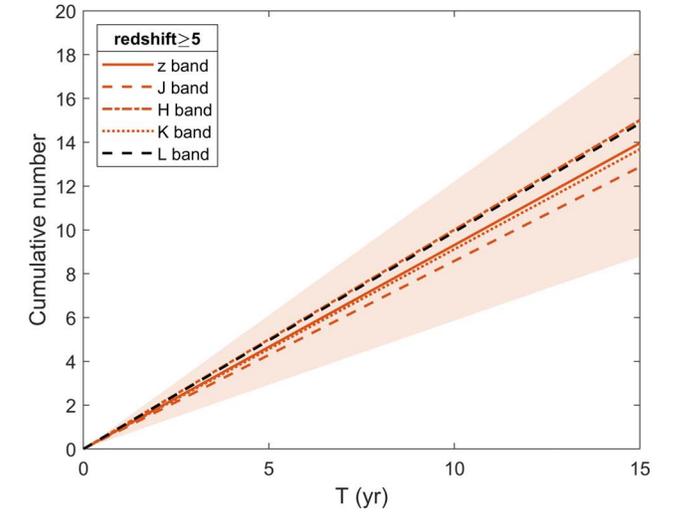

**Figure 2.** Predicted redshift distributions (top) and cumulative time distributions (bottom) of LGRB hosts in different bands, with limiting magnitude $m_{grizJHKL,\mathrm{lim}} = 28$ mag. The $x$-axis $T$ in the bottom panel represents the observational time of LGRBs. The legends are the same as in Figure 1.

**Figure 3.** The predicted number of low-mass (top; $M_* \leqslant 8\,M_\odot$) and high-redshift (bottom; $z \geqslant 5$) LGRB hosts accumulated over the observational time of LGRBs. The magnitude limit in each band is 28 mag and the legends are the same as in Figure 1.

$M_* \lesssim 10^8\,M_\odot$ challenging at redshift $\gtrsim 1$. Similar to the results of the IR bands, the detection of LGRB hosts in the $r$ and $i$ bands is moderate between the $g$ and $z$ bands.

Based on the above stellar mass limits, we compute the expected redshift distribution and cumulative time distribution of the LGRB hosts in each band. The distribution models of LGRB hosts have been mentioned in Section 2 and the LGRB energy in our predictions is limited by the fluence $S_{\mathrm{lim}} = 10^{-6}$ erg cm$^{-2}$. Considering the IGM absorption for galaxy light and the maximum measured redshift for LGRBs, we assume the maximum redshifts of LGRB hosts in the $grizJHKL$ bands are 3, 4, 5.5, 6.5, 10, 10, 10, and 10, respectively. The expected redshift distributions averaged over GRB observation time are presented in the top panel of Figure 2. Integrating the redshift distributions, we exhibit the predicted number of LGRB hosts accumulated over GRB observation time in the bottom panel of Figure 2. It is shown, with $m_{\lambda,\mathrm{lim}} = 28$ mag, that the $z$ band yields the highest peak of the expected redshift distributions. Consequently, although the assumed maximum redshifts in the IR bands are greater, the $z$ band still shows more strength in numbers for LGRB host observations. The detection expectation of LGRB hosts in the $g$ band is much lower than that of the other bands, probably because of the higher threshold of stellar mass in this band. Providing 15 yr Swift bright LGRBs, the $grizJHKL$ bands are projected to detect 116, 259, 277, 439, 266, 294, 274, and 316 LGRB hosts, respectively.

### 3.1. Low-mass and High-redshift LGRB Hosts

The typical nature of LGRB hosts is less luminous, indicating the potential of LGRBs to study low-mass galaxies (Savaglio et al. 2009; Trenti et al. 2012; Kelly et al. 2013; Greiner et al. 2015; Schulze et al. 2015; Perley et al. 2016a). As a sort of brightest event that can occur at high redshifts (Salvaterra et al. 2009a; Tanvir et al. 2009; Cucchiara et al. 2011), LGRBs are also expected to be used to constrain the properties of distant galaxies (Tanvir et al. 2012; Salvaterra 2015). In order to study the improvement extent of recent techniques for such science goals, we estimate the detection of low-mass and high-redshift LGRB hosts under next-generation conditions.

In the top panel of Figure 3, we show the predicted number of LGRB hosts with $M_* \leqslant 10^8 M_\odot$ in the *grizJHKL* bands. Supplying 15 yr LGRBs, there may exist 0.54, 31, 28, 143, 12, 20, 14, and 35 LGRB hosts with $M_* \leqslant 10^8 M_\odot$ in the *grizJHKL*-band observations with $m_{\lambda,\mathrm{lim}} = 28$ mag. The predicted results, especially in the *z* band, are largely ahead of the current detected number of low-mass hosts (Li & Zhang 2016), showing the great research potential of low-mass hosts supported by the next-generation telescopes.

The predictions of hosts with redshift $\geqslant 5$ in the *zJHKL*-band observations are presented in the bottom panel of Figure 3. The results of different bands are similar. For 15 yr LGRB observations, approximately 14 high-redshift hosts may be detected with $m_{zJHKL,\mathrm{lim}} = 28$ mag. This represents a notable improvement relative to the current number of distant LGRB hosts (Li & Zhang 2016; Perley et al. 2016b) and should enable better study of redshift evolution based on overall LGRB hosts. However, to achieve a statistically significant number for high-redshift hosts only, more locations of high-redshift LGRBs are still required.

To investigate the improvements provided by upcoming GRB detectors in the detection of distant hosts, we estimate the number of high-redshift host galaxies of LGRBs detected by the Space-based multi-band astronomical Variable Objects Monitor (SVOM; Wei et al. 2016; Atteia et al. 2022), which is planned to launch soon as a new GRB mission. The GRB trigger camera ECLAIRs (Godet et al. 2014) on board SVOM reaches a low-energy threshold of 4 keV. This allows SVOM/ECLAIRs to increase the sensitivity of high-redshift GRBs. In simulations, the quick-follow-up Visible Telescope (Fan et al. 2020) on board SVOM, which is capable of providing arcsecond localization, has the potential to detect afterglows of ~70% (Atteia et al. 2022) SVOM GRBs. This suggests that the access to LGRB hosts could be promising in the SVOM era. Since the proportion of optical afterglows of SVOM GRBs is similar to that of Swift GRBs, the fraction of good-localization LGRBs provided by Swift is still used here. With the field of view $\Delta\Omega = 2.0$ sr and the energy range 4–150 keV of SVOM/ECLAIRs, we find that the number of high-redshift hosts of SVOM LGRBs is 1.7 times that of the Swift LGRBs at $S \geqslant 10^{-6}$ erg cm$^{-2}$. For the *H* and *L* bands that yield the largest number of high-redshift LGRB hosts at $m_{HL,\mathrm{lim}} = 28$ mag, it would take ~17.6 yr to acquire a statistically significant number of distant hosts of SVOM LGRBs.

## 4. Discussion

In this work, the stellar mass–luminosity correlation is crucial for estimating the stellar mass limit, which is the basis for predicting LGRB host detection. However, as shown in Figure 5, this correlation is loose in the optical bands, maintaining uncertainties in our predicted results. To investigate if the less tight $M_*$–$M_\lambda(z)$ correlations in the optical bands will change our main conclusions, we compute the uncertainties contributed by the 1$\sigma$ confidence intervals of the $M_*$–$M_\lambda(z)$ correlations for our predictions in the *griz* bands, represented by corresponding colorful shadows in Figures 1–3. It is clear that the gap between the results of the *z* band and the *g* band is unable to be explained by the uncertainties of stellar mass–luminosity correlations. Although the middle dust extinction ($A_{V,\mathrm{average}} = 0.49$) of our sample may slightly aggravate the disadvantage of the *g* band, it should not be sufficient to overturn our predictions. We also compare the *r*-band results predicted by our method with the observations of Krühler et al. (2015; hereafter, K15), who measured emission-line spectroscopy of 96 host galaxies with 21 mag $\lesssim m_{R/r} \lesssim$ 26 mag at $0.1 <$ redshift $< 3.6$ through Very Large Telescope (VLT)/X-Shooter for 9 yr LGRBs. The majority of K15's LGRBs are contributed by Swift, with ~two-thirds of them having fluence $S \geqslant 10^{-6}$ erg cm$^{-2}$. Adopting the LGRB observation time and minimum magnitude of the K15 sample, our prediction suggests there are $76.49^{+27.61}_{-37.40}$ LGRB hosts in the *r*-band observation. This is in good agreement with the observed number of host galaxies of high-energy LGRBs in K15, considering the observable area of VLT covers about three-quarters of the sky (see Hjorth et al. 2012).

The stellar mass limit $M_{*,\mathrm{lim}}(z)$ is determined by the limiting magnitude of each band. While in previous sections, we assumed uniform limiting magnitudes for the *grizJHKL* bands, it is important to recognize that the sensitivities of these bands are typically different. Therefore, to address practical considerations, we estimate the potential detection of LGRB hosts across a sensitivity range of 23–30 mag and illustrate the results in Figure 4. This figure provides a direct representation of the expected number of host galaxies for Swift bright LGRBs as observed with specific limiting magnitudes in different bands. We also employ the sensitivities of JWST and CSST to estimate their capabilities of detecting LGRB host galaxies. For the JWST filters (F090W, F115W, F162M, F210M, and F360M), which fall within a wavelength range similar to the *zJHKL* bands, our predictions suggest the potential detection of 488, 314, 330, 310, and 331 LGRB hosts in 10 ks exposures,[6] when providing 15 yr Swift LGRBs with $S \geqslant 10^{-6}$ erg cm$^{-2}$. These include 190, 30, 38, 27, and 43 low-mass hosts and 19, 16, 18, 16, and 16 high-redshift hosts. Furthermore, in the context of optical band observations, we introduced the survey depth of the ultradeep field of CSST (Zhan 2021) to assess its extreme for LGRB host detection. With limiting magnitudes of 27.5, 27.2, 27.0, and 26.4 mag in the *griz* bands, we find that CSST may be able to detect 102, 202, 203, and 265 LGRB hosts, including 0.25, 12, 9, and 34 low-mass ones, for 15 yr Swift bright LGRBs.

We note that the final observed number of LGRB hosts also depends on the environment (dust, foreground light, and so on) along the line of sight and the artificial operations in the observations for different science purposes. These factors are challenging to quantify, thus the predicted results presented in this work, which only account for the properties of instruments, should be considered as upper limits of LGRB host detection. However, utilizing 70% operation time, CSST is planned to perform multicolor imaging and slitless spectrum surveys for galaxies within 17,500 deg$^2$ of the sky over a 10 yr period (all filters complete) with a fixed observation strategy. In this case, ~34, 64, 60, and 70 images and ~10 spectra covered by entire *griz* intervals could be approximately detected for host galaxies of decennial SVOM LGRBs. Since the Gamma-ray Monitor of SVOM can measure GRB spectra over a large energy range (15 keV–5 MeV), the connection between LGRBs and their hosts may be better understood in a wider channel with the collaboration between SVOM and CSST.

---

[6] https://jwst-docs.stsci.edu/jwst-near-infrared-camera/nircam-performance/nircam-sensitivity

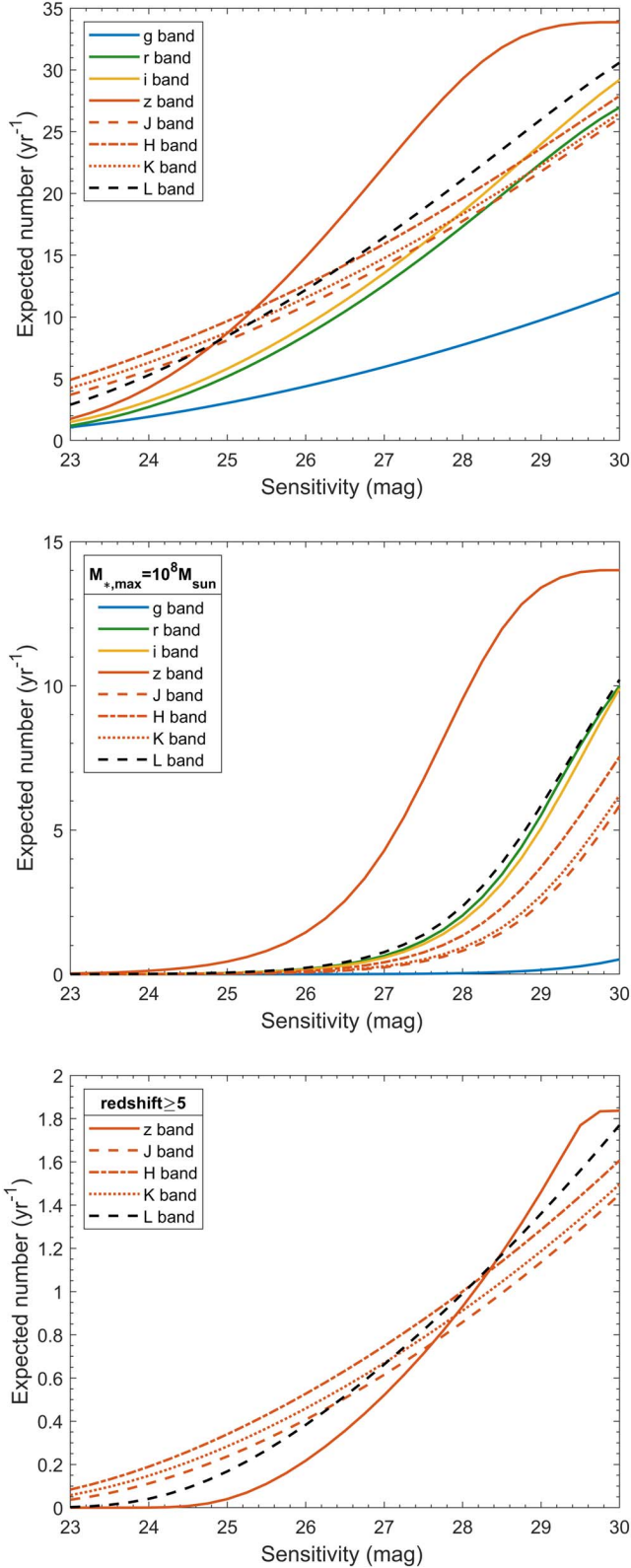

**Figure 4.** The predicted number of overall (top), low-mass (middle), and high-redshift (bottom) host galaxies of Swift LGRBs annualized by GRB observation time at different sensitivities. The legends are the same as in Figure 1.

## 5. Summary and Conclusion

Using the stellar mass function of host galaxies of LGRBs, we estimate the detection expectations of LGRB hosts in the $grizJHKL$ bands. By assuming the sensitivity $m_{\lambda,\text{lim}}$ of each band, we find that the $z$ band yields the largest number of LGRB hosts at $m_{\lambda,\text{lim}} \gtrsim 26$ mag. For the $riJHKL$ bands, it is sufficient that the limiting magnitudes improve by 2 mag to earn detection expectations similar to the $z$ band at middle and high sensitivities. Nonetheless, particularly at high sensitivities, such an extent of improvement seems insufficient to alter the disadvantage of the $g$ band, where the expected number is the least sensitive to the increase in limiting magnitudes. Since the $z$-band luminosity is useful for estimating the stellar masses of host galaxies, the advantage in the $z$ band provides a good opportunity to refine the stellar mass function of LGRB hosts (Lan et al. 2022).

The $z$ band is also the best interval for detecting host galaxies with $M_* \leqslant 10^8 M_\odot$. Given $m_{z,\text{lim}} \sim 26.5$ mag, the detectable number of low-mass hosts could be statistically significant, and could potentially reach $\sim 10^2$ if the limiting magnitude is to increase to $m_{z,\text{lim}} \sim 27.5$ mag, for 15 yr Swift LGRBs with $S \geqslant 10^{-6}$ erg cm$^{-2}$. This shows that enhancing the depth of LGRB host surveys in the $z$ band to JWST's capability may substantially improve the statistical study of low-mass hosts. Additionally, the investigation of low-mass hosts could further contribute to a better understanding of the formation environments of LGRBs. Although LGRBs have a preference for occurring in metal-poor environments, a few instances have been discovered to be associated with high-metallicity hosts (see K15; Perley et al. 2016b and references therein). Kelly et al. (2013) indicated that the host of an LGRB that exploded near a metal-rich galaxy may be a low-metallicity dwarf galaxy of insufficient luminosity to be easily detected. This finding may partially account for the positioning coincidence between high-metallicity galaxies and some LGRBs. However, an increased number of low-luminosity LGRB hosts are required to bolster the evidence. With the survey of LGRB hosts going deeper, the formation environments of LGRBs can be clearer.

The optimal bands for detecting high-redshift LGRB hosts differ between low and high sensitivities. While the $z$ band remains the most promising in terms of high sensitivities, its efficiency diminishes significantly at lower sensitivities. On the other hand, the $L$ band, due to its detection expectation of distant hosts being similar to that of the optimal band at any sensitivity, emerges as a preferable option. However, it is necessary to achieve a magnitude limit of $m_{L,\text{lim}} \gtrsim 30$ mag in order to obtain a statistically significant number of high-redshift LGRB hosts for 15 yr Swift LGRBs. This is mainly due to the paucity of the observed number of high-redshift LGRBs in the Swift era. To efficiently increase the detectable number of distant LGRB hosts, a preference for the detection and localization of high-redshift GRBs is needed for future GRB detections. For this aim, the implementation of a low-energy GRB trigger camera capable of observing faint GRBs, alongside the establishment of rapid near-IR follow-up telescopes, to access the afterglows of high-redshift LGRBs, is suggested by the study of Ghirlanda et al. (2015).

## Acknowledgments

We are very grateful to Bing-Xiao Xu and Wen-Jin Xie for their valuable discussions that improved this work. This work

Table 1
Observational Parameters of LGRB Hosts with $A_V \lesssim 1$

| GRBs | Redshift | log $M_*$ ($M_\odot$) | $A_V$ | g | r | i | z | References | J | H | K | References |
|---|---|---|---|---|---|---|---|---|---|---|---|---|
| GRB 990712 | 0.4340 | 9.29 | 0.39 | ⋯ | ⋯ | ⋯ | ⋯ | ⋯ | 21.68 | 21.60 | 21.85 | Savaglio et al. (2009) |
| GRB 011121 | 0.3620 | 9.81 | 0.38 | ⋯ | ⋯ | ⋯ | ⋯ | ⋯ | 20.47 | ⋯ | ⋯ | Savaglio et al. (2009) |
| GRB 030329 | 0.1680 | 7.74 | <0.10 | ⋯ | ⋯ | ⋯ | ⋯ | ⋯ | 22.40 | 22.54 | ⋯ | Savaglio et al. (2009) |
| GRB 031203 | 0.1055 | 8.82 | 0.03 | ⋯ | ⋯ | ⋯ | ⋯ | ⋯ | ⋯ | 18.53 | 18.03 | Savaglio et al. (2009) |
| GRB 050219 | 0.2110 | 9.98 | 0.00[a] | 20.85 | 19.76 | 19.38 | 19.18 | Rossi et al. (2014) | 18.52 | 18.30 | 18.44 | Rossi et al. (2014) |
| GRB 051008 | 2.9000 | 9.69 | 0.85 | 24.46 | ⋯ | ⋯ | 24.15 | Perley et al. (2013) | ⋯ | ⋯ | ⋯ | ⋯ |
| GRB 051022 | 0.8070 | 10.42 | 0.73 | ⋯ | 21.93 | 21.44 | 21.01 | Perley et al. (2013) | ⋯ | ⋯ | ⋯ | ⋯ |
| GRB 060202 | 0.7850 | 9.04 | 1.00 | 24.45 | ⋯ | ⋯ | ⋯ | Perley et al. (2013) | ⋯ | ⋯ | ⋯ | ⋯ |
| GRB 060319 | 1.1720 | 10.33 | 0.91 | 24.51 | ⋯ | ⋯ | 22.91 | Perley et al. (2013) | ⋯ | ⋯ | ⋯ | ⋯ |
| GRB 061121 | 1.3145 | 10.20 | 0.45 | 22.95 | ⋯ | 22.66 | 22.33 | Perley et al. (2015) | ⋯ | ⋯ | ⋯ | ⋯ |
| GRB 061222 | 2.0880 | 8.04 | 0.00 | ⋯ | ⋯ | ⋯ | 25.55 | Perley et al. (2013) | ⋯ | ⋯ | ⋯ | ⋯ |
| GRB 070306 | 1.4959 | 10.36 | 0.21[b] | 22.90 | 23.08 | 22.81 | 22.86 | Krühler et al. (2011) | 21.62 | 21.20 | 21.38 | Krühler et al. (2011) |
| GRB 070802 | 2.4541 | 9.85 | ∼1.00 | ⋯ | ⋯ | 25.50 | ⋯ | Krühler et al. (2011) | 24.50 | ⋯ | 23.40 | Krühler et al. (2011) |
| GRB 080517 | 0.0890 | 9.60 | 0.16 | 18.03 | 17.73 | 17.46 | ⋯ | Stanway et al. (2015) | 17.37 | 17.22 | 17.50 | Stanway et al. (2015) |
| GRB 080605 | 1.6403 | 9.90 | 0.20 | 23.15 | 22.82 | 22.81 | 22.76 | Krühler et al. (2011) | 21.90 | 22.30 | ⋯ | Krühler et al. (2011) |
| GRB 080805 | 1.5042 | 9.70 | 1.01[c] | ⋯ | ⋯ | 25.70 | ⋯ | Krühler et al. (2011) | 23.60 | ⋯ | 23.10 | Krühler et al. (2011) |
| GRB 081109 | 0.9790 | 9.82 | 1.00 | 23.07 | 22.74 | 22.01 | 22.01 | Krühler et al. (2011) | 21.40 | 21.50 | 21.05 | Krühler et al. (2011) |
| GRB 090323 | 3.5690 | 11.20 | 0.14 | ⋯ | 24.87 | 24.25 | ⋯ | McBreen et al. (2010) | ⋯ | ⋯ | ⋯ | ⋯ |
| GRB 090417B | 0.3450 | 10.14 | 0.54[d] | 22.02 | 21.62 | 21.41 | 20.78 | Holland et al. (2010) | ⋯ | ⋯ | ⋯ | ⋯ |
| GRB 091127 | 0.4903 | 8.60 | <0.50[e] | 24.23 | 23.15 | 22.86 | 22.39 | Vergani et al. (2011) | ⋯ | ⋯ | ⋯ | ⋯ |
| GRB 100621 | 0.5420 | 8.98 | 0.60 | 21.86 | 21.48 | 21.15 | 21.46 | Krühler et al. (2011) | 21.43 | 21.18 | 21.23 | Krühler et al. (2011) |
| GRB 120624B | 2.1974 | 10.60 | 1.00 | 24.80 | 24.17 | 24.15 | 23.89 | de Ugarte Postigo et al. (2013) | 22.78 | 22.24 | 21.39 | de Ugarte Postigo et al. (2013) |
| GRB 130427 | 0.3399 | 9.32 | 0.05 | 21.98 | 21.26 | 21.19 | 21.01 | Perley et al. (2014) | 20.84 | 20.73 | 20.74 | Perley et al. (2014) |
| GRB 130702 | 0.1450 | 7.90 | 0.00 | ⋯ | 23.01 | ⋯ | ⋯ | Kelly et al. (2013) | ⋯ | ⋯ | ⋯ | ⋯ |
| GRB 140506 | 0.8891 | 9.00 | ∼1.00 | ⋯ | 24.43 | 23.70 | 23.71 | Fynbo et al. (2014) | ⋯ | ⋯ | ⋯ | ⋯ |
| GRB 161108 | 1.1590 | 11.05 | 0.30 | 23.46 | 22.70 | 22.18 | 21.99 | Chrimes et al. (2018) | 17.19 | 16.34 | 16.64 | Chrimes et al. (2018) |

**Note.** The magnitudes offered by Perley et al. (2013) and Perley et al. (2015) are the Sloan Digital Sky Survey magnitudes that are close to the AB results. The other magnitudes are reported in the AB system. The references of $A_V$ with a superscript refer to the following: (a) Hunt et al. (2014); (b) Perley et al. (2015); (c) Greiner et al. (2011); (d) Hunt et al. (2014); and (e) Cobb et al. (2010).

is partially supported by the Natural Science Foundation of China (Nos. 11773003, U1931201, 12041306, 12103089, and 12321003), the China Manned Space Project (Nos. CMS-CSST-2021-B11 and CMS-CSST-2021-A12), the National Key Research and Development Program of China (2022SKA0130100), the Natural Science Foundation of Jiangsu Province (grant No. BK20211000), the International Partnership Program of the Chinese Academy of Sciences for Grand Challenges (114332KYSB20210018), and the Major Science and Technology Project of Qinghai Province (2019-ZJ-A10).

## Appendix
## $M_*$–$M_\lambda$ Correlations in Different Bands

GHostS is a public database that offers multiple observational properties of GRB hosts. Based on this database, we collect the stellar masses and apparent magnitudes of LGRB hosts in the *grizJHK* bands to empirically estimate their stellar mass–luminosity correlations ($M_*$–$M_\lambda(z)$, where $\lambda$ is the band for host observation). As the apparent magnitude measured in the optical band is strongly affected by dust extinction, here we only take into account LGRB hosts with $A_V \lesssim 1$ to reduce the uncertainties of the $M_*$–$M_\lambda(z)$ correlations. The relevant properties of LGRB hosts are presented in Table 1. Throughout this paper, the absolute magnitude in the rest frame is calculated by the following formula (Blanton & Roweis 2007; Perley et al. 2016b):

$$M_\lambda(z) = m_{\lambda,\text{obs}} - \mu(z) + 2.5\log(1+z), \quad \text{(A1)}$$

where $\mu(z)$ is the distance modulus at redshift $z$. Finally, we empirically use the linear formula $\log M_* = wM_\lambda(z) + k$ to fit the $M_*$–$M_\lambda(z)$ correlations. The results are shown in Figure 5 and the best-fitting correlations in different bands can be expressed as (see the *L*-band result in Lan et al. 2022):

$$\begin{aligned}
\log M_* &= -0.33 M_g + 3.39; \\
\log M_* &= -0.54 M_r - 1.02; \\
\log M_* &= -0.50 M_i - 0.38; \\
\log M_* &= -0.74 M_z - 5.66; \\
\log M_* &= -0.34 M_J + 2.38; \\
\log M_* &= -0.33 M_H + 2.41; \\
\log M_* &= -0.33 M_K + 2.51.
\end{aligned} \quad \text{(A2)}$$

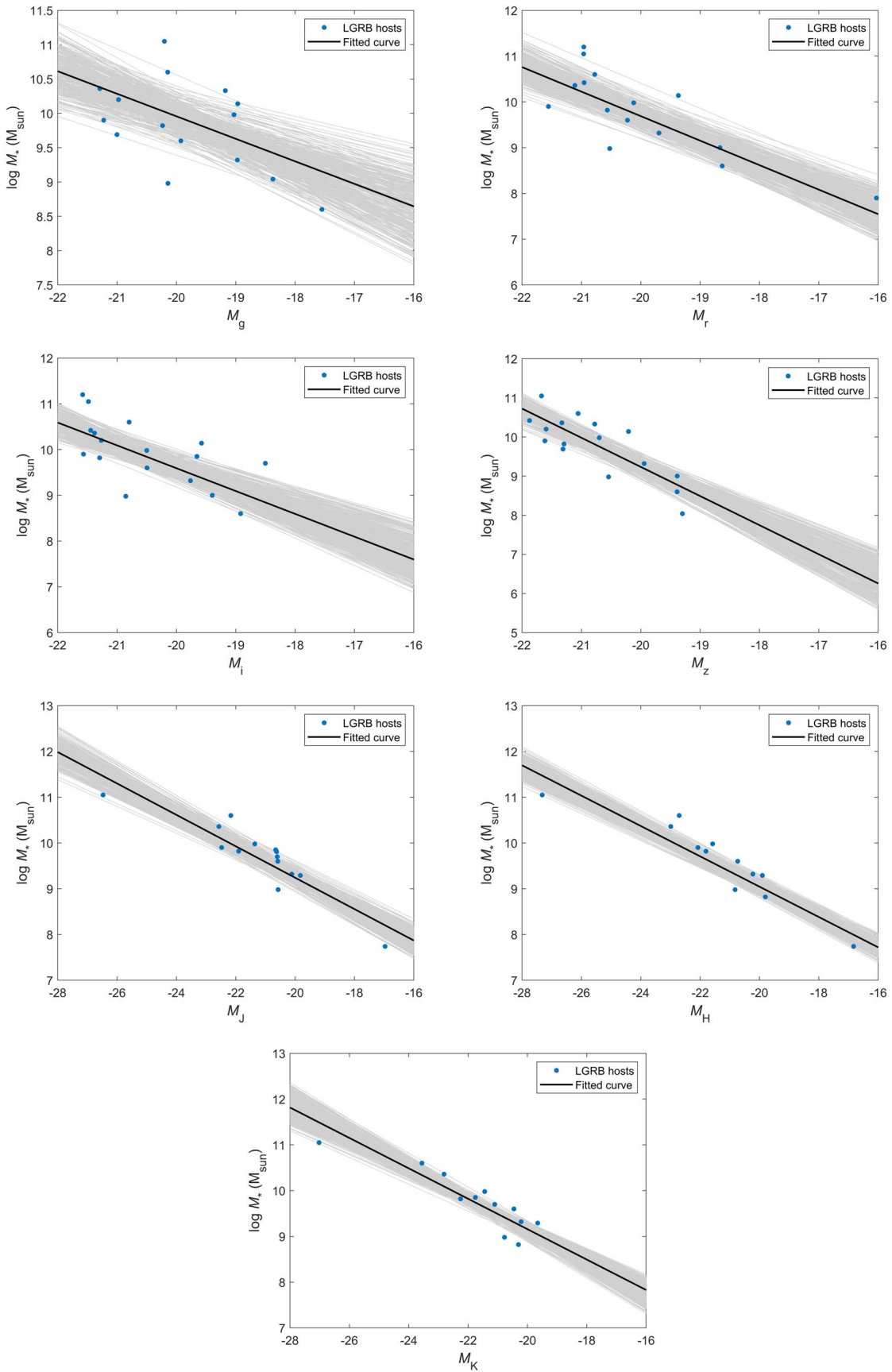

**Figure 5.** The stellar mass–luminosity ($M_*$–$M_\lambda$) correlations of LGRB hosts in the *grizJHK* bands. The observational band is shown in the subscript $\lambda$ and the shaded regions are $1\sigma$ confidence intervals of the corresponding correlations.

**References**

Abbott, B. P., Abbott, R., Abbott, T. D., et al. 2017, PhRvL, 119, 161101
Abdo, A. A., Ackermann, M., Ajello, M., et al. 2010, ApJ, 720, 435
Ahumada, T., Singer, L. P., Anand, S., et al. 2021, NatAs, 5, 917
Ajello, M., Costamante, L., Sambruna, R. M., et al. 2009, ApJ, 699, 603
Ajello, M., Shaw, M. S., Romani, R. W., et al. 2012, ApJ, 751, 108
Amati, L., Frontera, F., Tavani, M., et al. 2002, A&A, 390, 81
Antonelli, L. A., D'Avanzo, P., Perna, R., et al. 2009, A&A, 507, L45
Atteia, J. L., Cordier, B., & Wei, J. 2022, IJMPD, 31, 2230008
Band, D., Matteson, J., Ford, L., et al. 1993, ApJ, 413, 281
Basa, S., Cuby, J. G., Savaglio, S., et al. 2012, A&A, 542, A103
Berger, E. 2009, ApJ, 690, 231
Berger, E. 2014, ARA&A, 52, 43
Blanchard, P. K., Berger, E., & Fong, W.-f. 2016, ApJ, 817, 144
Blanton, M. R., & Roweis, S. 2007, AJ, 133, 734
Bloom, J. S., Kulkarni, S. R., & Djorgovski, S. G. 2002, AJ, 123, 1111
Chrimes, A. A., Stanway, E. R., Levan, A. J., et al. 2018, MNRAS, 478, 2
Cobb, B. E., Bloom, J. S., Perley, D. A., et al. 2010, ApJL, 718, L150
Cucchiara, A., Levan, A. J., Fox, D. B., et al. 2011, ApJ, 736, 7
de Ugarte Postigo, A., Campana, S., Thöne, C. C., et al. 2013, A&A, 557, L18
Drory, N., & Alvarez, M. 2008, ApJ, 680, 41
Fan, X., Zou, G., Qiu, Y., et al. 2020, ApOpt, 59, 3049
Firmani, C., Avila-Reese, V., Ghisellini, G., & Tutukov, A. V. 2004, ApJ, 611, 1033
Fong, W., & Berger, E. 2013, ApJ, 776, 18
Fruchter, A. S., Levan, A. J., Strolger, L., et al. 2006, Natur, 441, 463
Fynbo, J. P. U., Krühler, T., Leighly, K., et al. 2014, A&A, 572, A12
Gehrels, N., Norris, J. P., Barthelmy, S. D., et al. 2006, Natur, 444, 1044
Ghirlanda, G., & Salvaterra, R. 2022, ApJ, 932, 10
Ghirlanda, G., Salvaterra, R., Ghisellini, G., et al. 2015, MNRAS, 448, 2514
Godet, O., Nasser, G., Atteia, J., et al. 2014, Proc. SPIE, 9144, 914424
Goldstein, A., Veres, P., Burns, E., et al. 2017, ApJL, 848, L14
Graham, J. F., & Fruchter, A. S. 2013, ApJ, 774, 119
Greiner, J., Fox, D. B., Schady, P., et al. 2015, ApJ, 809, 76
Greiner, J., Krühler, T., Klose, S., et al. 2011, A&A, 526, A30
Hjorth, J., Malesani, D., Jakobsson, P., et al. 2012, ApJ, 756, 187
Holland, S. T., Sbarufatti, B., Shen, R., et al. 2010, ApJ, 717, 223
Hunt, L. K., Palazzi, E., Michałowski, M. J., et al. 2014, A&A, 565, A112
Japelj, J., Vergani, S. D., Salvaterra, R., et al. 2018, A&A, 617, A105
Kaneko, Y., Preece, R. D., Briggs, M. S., et al. 2006, ApJS, 166, 298
Kelly, P. L., Filippenko, A. V., Fox, O. D., Zheng, W., & Clubb, K. I. 2013, ApJL, 775, L5
Kelly, P. L., Filippenko, A. V., Modjaz, M., & Kocevski, D. 2014, ApJ, 789, 23
Kouveliotou, C., Meegan, C. A., Fishman, G. J., et al. 1993, ApJL, 413, L101
Krühler, T., Greiner, J., Schady, P., et al. 2011, A&A, 534, A108
Krühler, T., Malesani, D., Fynbo, J. P. U., et al. 2015, A&A, 581, A125
Lan, G.-X., Wei, J.-J., Li, Y., Zeng, H.-D., & Wu, X.-F. 2022, ApJ, 938, 129
Lan, G.-X., Wei, J.-J., Zeng, H.-D., Li, Y., & Wu, X.-F. 2021, MNRAS, 508, 52
Lan, G.-X., Zeng, H.-D., Wei, J.-J., & Wu, X.-F. 2019, MNRAS, 488, 4607
Langer, N., & Norman, C. A. 2006, ApJL, 638, L63
Le, T., & Dermer, C. D. 2007, ApJ, 661, 394
Levan, A., Gompertz, B. P., Salafia, O. S., et al. 2023, arXiv:2307.02098
Levesque, E. M., Bloom, J. S., Butler, N. R., et al. 2010a, MNRAS, 401, 963
Levesque, E. M., Kewley, L. J., Berger, E., & Zahid, H. J. 2010b, AJ, 140, 1557
Li, L.-X. 2008, MNRAS, 388, 1487
Li, Y., Zhang, B., & Lü, H.-J. 2016, ApJS, 227, 7
Lloyd-Ronning, N. M., Fryer, C. L., & Ramirez-Ruiz, E. 2002, ApJ, 574, 554
McBreen, S., Krühler, T., Rau, A., et al. 2010, A&A, 516, A71
Modjaz, M., Kewley, L., Kirshner, R. P., et al. 2008, AJ, 135, 1136
Nava, L., Salvaterra, R., Ghirlanda, G., et al. 2012, MNRAS, 421, 1256
Perley, D. A., Cenko, S. B., Corsi, A., et al. 2014, ApJ, 781, 37
Perley, D. A., Krühler, T., Schulze, S., et al. 2016a, ApJ, 817, 7
Perley, D. A., Levan, A. J., Tanvir, N. R., et al. 2013, ApJ, 778, 128
Perley, D. A., Perley, R. A., Hjorth, J., et al. 2015, ApJ, 801, 102
Perley, D. A., Tanvir, N. R., Hjorth, J., et al. 2016b, ApJ, 817, 8
Pescalli, A., Ghirlanda, G., Salvaterra, R., et al. 2016, A&A, 587, A40
Preece, R. D., Briggs, M. S., Mallozzi, R. S., et al. 2000, ApJS, 126, 19
Rastinejad, J. C., Gompertz, B. P., Levan, A. J., et al. 2022, Natur, 612, 223
Robertson, B. E., & Ellis, R. S. 2012, ApJ, 744, 95
Rossi, A., Piranomonte, S., Savaglio, S., et al. 2014, A&A, 572, A47
Rossi, A., Rothberg, B., Palazzi, E., et al. 2022, ApJ, 932, 1
Salvaterra, R. 2015, JHEAp, 7, 35
Salvaterra, R., Campana, S., Vergani, S. D., et al. 2012, ApJ, 749, 68
Salvaterra, R., Della Valle, M., Campana, S., et al. 2009a, Natur, 461, 1258
Salvaterra, R., Guidorzi, C., Campana, S., Chincarini, G., & Tagliaferri, G. 2009b, MNRAS, 396, 299
Savaglio, S., Glazebrook, K., & Le Borgne, D. 2009, ApJ, 691, 182
Schulze, S., Chapman, R., Hjorth, J., et al. 2015, ApJ, 808, 73
Stanek, K. Z., Gnedin, O. Y., Beacom, J. F., et al. 2006, AcA, 56, 333
Stanway, E. R., Levan, A. J., Tanvir, N., et al. 2015, MNRAS, 446, 3911
Sun, H., Wang, C. W., Yang, J., et al. 2023, arXiv:2307.05689
Svensson, K. M., Levan, A. J., Tanvir, N. R., Fruchter, A. S., & Strolger, L. G. 2010, MNRAS, 405, 57
Tanvir, N. R., Fox, D. B., Levan, A. J., et al. 2009, Natur, 461, 1254
Tanvir, N. R., Levan, A. J., Fruchter, A. S., et al. 2012, ApJ, 754, 46
Trenti, M., Perna, R., Levesque, E. M., Shull, J. M., & Stocke, J. T. 2012, ApJL, 749, L38
Troja, E., Fryer, C. L., O'Connor, B., et al. 2022, Natur, 612, 228
Vergani, S. D., Flores, H., Covino, S., et al. 2011, A&A, 535, A127
Vergani, S. D., Salvaterra, R., Japelj, J., et al. 2015, A&A, 581, A102
Wanderman, D., & Piran, T. 2010, MNRAS, 406, 1944
Wang, F. Y., & Dai, Z. G. 2011, ApJL, 727, L34
Wang, F. Y., & Dai, Z. G. 2014, ApJS, 213, 15
Wei, J., Cordier, B., Antier, S., et al. 2016, arXiv:1610.06892
Wei, J.-J., & Wu, X.-F. 2017, IJMPD, 26, 1730002
Wei, J.-J., Wu, X.-F., Melia, F., Wei, D.-M., & Feng, L.-L. 2014, MNRAS, 439, 3329
Woosley, S. E., & Bloom, J. S. 2006, ARA&A, 44, 507
Yang, J., Ai, S., Zhang, B.-B., et al. 2022, Natur, 612, 232
Zeng, H., Melia, F., & Zhang, L. 2016, MNRAS, 462, 3094
Zeng, H., Yan, D., & Zhang, L. 2014, MNRAS, 441, 1760
Zhan, H. 2021, ChSBu, 66, 1290
Zhang, B. B., Liu, Z. K., Peng, Z. K., et al. 2021, NatAs, 5, 911